\documentclass[aps,prd,twocolumn,superscriptaddress,nofootinbib,preprintnumbers]{revtex4-1}
\usepackage{amsmath}
\usepackage{graphicx}
\usepackage{subfigure}
\usepackage{amssymb}
\usepackage{xcolor}
\usepackage{braket}
\usepackage{multirow}
\usepackage{cancel}
\usepackage{color}
\usepackage{ulem}
\usepackage{listings}
\usepackage{xcolor}
\usepackage{float}
\usepackage{slashed}

\usepackage{wrapfig}
\usepackage{mathtools}

\usepackage{tikz}
\usetikzlibrary{arrows,shapes}
\usetikzlibrary{trees}
\usetikzlibrary{matrix,arrows} 				
\usetikzlibrary{positioning}				
\usetikzlibrary{calc,through}				
\usetikzlibrary{decorations.pathreplacing}  
\usepackage{pgffor}							

\usetikzlibrary{decorations.pathmorphing}	
\usetikzlibrary{decorations.markings}
\tikzset{
    vector/.style={decorate, decoration={snake}, draw},
	provector/.style={decorate, decoration={snake,amplitude=2.5pt}, draw},
	antivector/.style={decorate, decoration={snake,amplitude=-2.5pt}, draw},
    fermion/.style={draw=black, postaction={decorate},
        decoration={markings,mark=at position .55 with {\arrow[draw=black]{>}}}},
    fermionbar/.style={draw=black, postaction={decorate},
        decoration={markings,mark=at position .55 with {\arrow[draw=black]{<}}}},
    fermionnoarrow/.style={draw=black},
    gluon/.style={decorate, draw=black,
        decoration={coil,amplitude=4pt, segment length=5pt}},
    scalar/.style={dashed,draw=black, postaction={decorate},
        decoration={markings,mark=at position .55 with {\arrow[draw=black]{>}}}},
    scalarbar/.style={dashed,draw=black, postaction={decorate},
        decoration={markings,mark=at position .55 with {\arrow[draw=black]{<}}}},
    scalarnoarrow/.style={dashed,draw=black},
    electron/.style={draw=black, postaction={decorate},
        decoration={markings,mark=at position .55 with {\arrow[draw=black]{>}}}},
	bigvector/.style={decorate, decoration={snake,amplitude=4pt}, draw},
}

\tikzstyle{block} = [draw, rectangle, 
    minimum height=3em, minimum width=6em]

\usepackage[colorlinks,citecolor=blue]{hyperref}
\usepackage{amsmath}
\usepackage{wrapfig}

\usepackage[T1]{fontenc} 
\usepackage[utf8]{inputenc} 
\usepackage{times}

\newcommand{\be}{\begin{equation}}
\newcommand{\ee}{\end{equation}}
\newcommand{\beq}{\begin{equation}}
\newcommand{\eeq}{\end{equation}}
\newcommand{\bea}{\begin{eqnarray}}
\newcommand{\eea}{\end{eqnarray}}
\newcommand{\besp}{\begin{equation}\begin{split}}
\newcommand{\eesp}{\end{split}\end{equation}}

\newcommand{\nn}{\nonumber}

\newcommand{\Dfbd}{\mathord{\buildrel{\lower3pt\hbox{$\scriptscriptstyle\leftrightarrow$}}\over {D}_{\mu}}}
\newcommand{\ave}[1]{\left\langle #1\right\rangle}

\def\0{\textbf{0}}
\def\1{\textbf{1}}
\def\2{\textbf{2}}
\def\3{\textbf{3}}
\def\4{\textbf{4}}
\def\5{\textbf{5}}
\def\6{\textbf{6}}
\def\7{\textbf{7}}
\def\8{\textbf{8}}
\def\9{\textbf{9}}

\def\d{\text{d}}

\usepackage{fontawesome5}

\begin{document}

\title{Can we live in a baby universe formed by a delayed first-order phase transition?}

\author{Qing-Hong Cao}
\email{qinghongcao@pku.edu.cn}
\affiliation{School of Physics, Peking University, Beijing 100871, China}
\affiliation{Center for High Energy Physics, Peking University, Beijing 100871, China}
\affiliation{School of Physics, Zhengzhou University, Zhengzhou 450001, China}

\author{Masanori Tanaka}
\email{tanaka@pku.edu.cn}
\affiliation{Center for High Energy Physics, Peking University, Beijing 100871, China}

\author{Jun-Chen~Wang}
\email{junchenwang@stu.pku.edu.cn}
\affiliation{School of Physics, Peking University, Beijing 100871, China}

\author{Ke-Pan Xie}
\email{kpxie@buaa.edu.cn}
\affiliation{School of Physics, Beihang University, Beijing 100191, P. R. China}

\author{Jing-Jun Zhang}
\email{zhang\_jingjun@stu.pku.edu.cn}
\affiliation{School of Physics, Peking University, Beijing 100871, China}

\preprint{CPTNP-2025-014}

\begin{abstract}
We examine the idea that our universe began as a baby universe and show that this is feasible in a gauged $U(1)_{B-L}$ extension of the Standard Model with the classically conformal principle. For the first time, we define a measure to describe the probability that we reside in a baby universe, and find that it can be close to 1 in a considerable portion of the parameter space. The framework is consistent with current cosmological data, and it predicts the existence of a heavy neutral gauge boson, which could be detected at colliders, thereby offering a direct link between early-universe dynamics and experimentally testable signatures at the TeV scale.
\end{abstract}
\maketitle

\textbf{Introduction}. The concept of multiple universes, called the multiverse, has been proposed to explain why our universe is so well organized~\cite{Hogan:1999wh, Linde:2015edk}. One realization of the multiverse scenario is the baby universes, which can emerge during a cosmic first-order phase transition (FOPT), where stochastic bubble nucleation leaves rare regions in the false vacuum domain (FVD). These regions subsequently evolve into inflating baby universes, appearing as primordial black holes (PBHs), which is called the super-critical PBH~\cite{Garriga:2015fdk}, to observers in the parent universe~\cite{Sato:1981bf,Kodama:1981gu,Kodama:1982sf,Maeda:1981gw,Berezin:1983dz,Berezin:1982ur,Blau:1986cw,Berezin:1987bc,Deng:2016vzb,Deng:2017uwc,Kusenko:2020pcg,Gouttenoire:2023gbn,Gouttenoire:2025ofv,Hashino:2025fse}. Such processes are particularly enhanced in slow or supercooled FOPTs. Usually, the baby universes born in FOPTs are in an eternal inflationary period. Hence, it is taken for granted that we live in the parent universe. However, this work demonstrates that such an eternal inflation can be naturally terminated by low-scale QCD phase transitions within a classically conformal (CC) scenario. In this case, standard cosmology could be restored, and we may live in a baby universe.

The CC principle is applied to the scalar sector of the Standard Model (SM) to address the hierarchy problem~\cite{Bardeen:1995kv,Meissner:2007xv}. In this framework, the tree-level scalar potential is conformal, containing no bare mass term. Loop corrections generate a logarithmic potential that breaks conformal invariance, triggering the symmetry breaking. The flatness of the CC potential plays a dual role in the early universe. First, the thermal mass term always creates a local minimum at the origin of field space, forming a false vacuum that can trap the universe, leading to supercooling and the possible formation of a baby universe. Second, as the universe cools, the barrier near the false vacuum diminishes, making it highly sensitive to low-scale interactions. Consequently, a QCD phase transition can significantly alter the vacuum structure of a baby universe, allowing it to escape eternal inflation.

\begin{figure}
\centering
\includegraphics[width=0.4\textwidth]{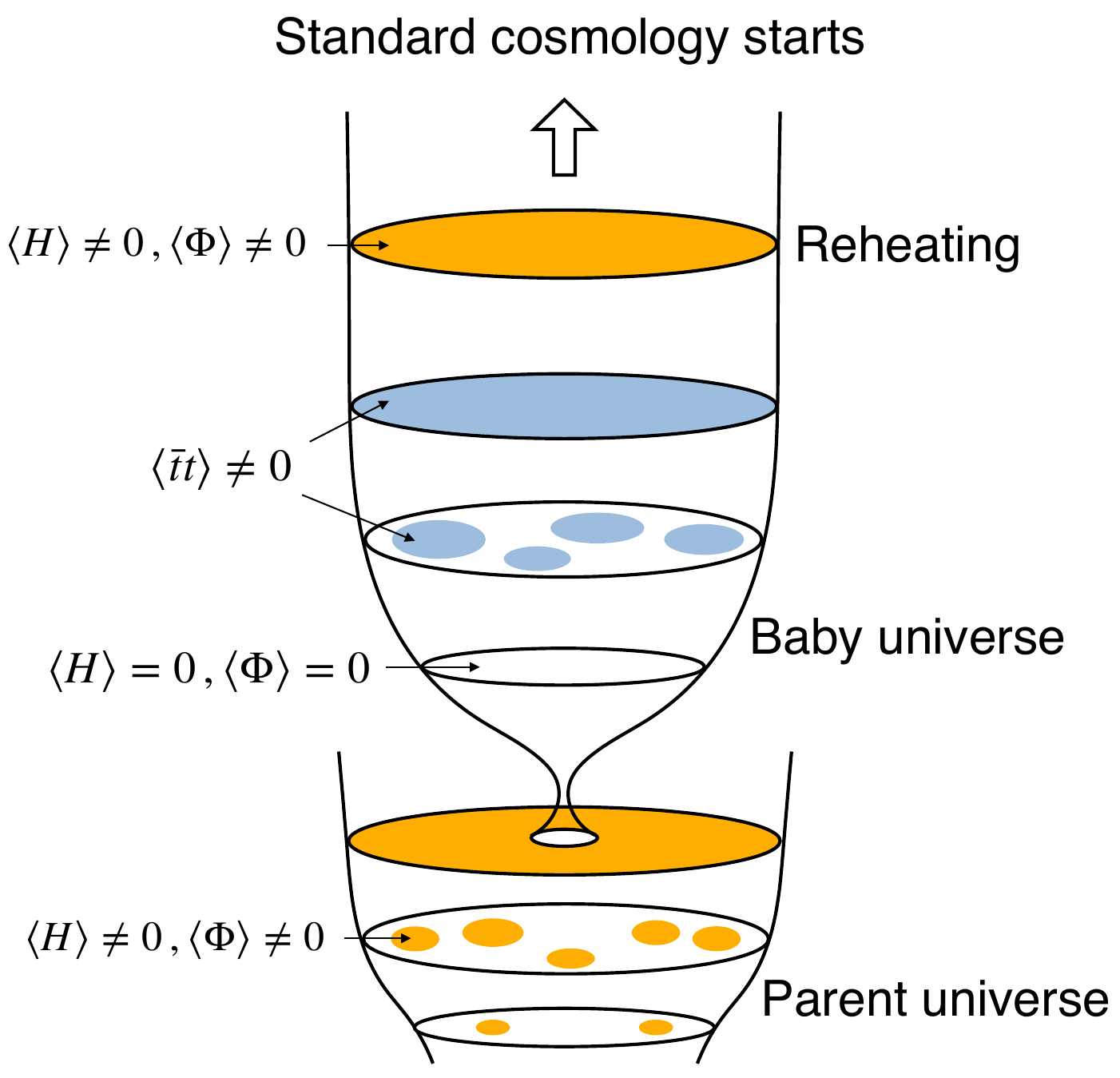}
\caption{Sketch of our scenario. In the parent universe, some rare regions remain in the false vacuum (white) until surrounded by true vacuum regions (orange). Those regions could collapse into super-critical PBHs, inside which an inflating baby universe is born. At a late time, the QCD phase transition (blue) triggers the $B-L$ and EW symmetry breaking (orange), starting the standard cosmology.
\label{fig:history}}
\end{figure}

In this work, we demonstrate the viability of this scenario and, for the first time, introduce a measure to quantify the probability that we live in a baby universe. As an example, we focus on the minimal $B-L$ model~\cite{Iso:2009ss,Iso:2009nw,Chun:2013soa}, which not only relaxes the hierarchy problem but also explains the neutrino mass via the type-I seesaw mechanism~\cite{Minkowski:1977sc,Yanagida:1979as}. In this model, the baryon-minus-lepton number is gauged, resulting in a new gauge boson $Z'$. Three generations of right-handed neutrinos $\nu_R$ with $B-L$ charge $-1$ are introduced to cancel gauge anomalies. Additionally, a SM singlet scalar $\Phi = (\varphi + i\eta)/\sqrt{2}$ with $B-L$ charge $+2$ is included to break $U(1)_{B-L}$ simultaneously. Under the CC principle, the joint potential of $\Phi$ and the SM Higgs doublet $H=\left(G^+,(h+iG^0)/\sqrt{2}\right)^T$ realizes a unique cosmological history, as depicted in Fig.~\ref{fig:history}, leading to the formation of baby universes. The details are elaborated below.

\textbf{The scalar potential}. At tree-level,
\begin{align}
\label{eq:Vtree}
V_{\rm tree}(H,\Phi)=\lambda_h|H|^4+\lambda_\varphi|\Phi|^4+\lambda_{h\varphi}|H|^2|\Phi|^2
\end{align}
is scale-invariant, and it preserves both the $U(1)_{B-L}$ and electroweak (EW) symmetries. However, both the conformal and gauge symmetries are broken by the radiative Coleman-Weinberg corrections~\cite{Coleman:1973jx}. Assuming that the $U(1)_{B-L}$ scale is much higher than the EW scale, we can use the sequential symmetry breaking approximation~\cite{Chataignier:2018kay}, in which the potential including the one-loop correction is
\begin{align}
V_{1\ell}(\varphi) \approx \frac{3 \overline{m}_{Z'}^4}{64 \pi^2} \left( \ln \frac{\varphi^2}{v_{\varphi}^2} - \frac{1}{2} \right) \,,
\end{align}
with $\overline{m}_{Z'}=2g_{B-L}\varphi$, which yields a VEV $\ave{\varphi}=v_\varphi$, breaking the $U(1)_{B-L}$ spontaneously, providing masses for $Z'$ and $\nu_R$. This induces a mass squared term $\lambda_{h\varphi}v_\varphi^2|H|^2/2$ along the Higgs direction, breaking the EW symmetry via a negative $\lambda_{h\varphi}\approx-m_h^2/v_\varphi^2$, where $m_h\approx125$ GeV is the Higgs boson mass.

The scalar potential at finite temperature along the $B-L$ scalar direction is given by
\begin{align}
\label{eq:VT}
V_{\rm tot}(\varphi, T) = V_{1\ell}(\varphi) + V_{T}(\varphi, T) \,,
\end{align}
where the one-loop thermal correction is dominated by the $Z'$ contribution, i.e.
\begin{align}
V_{T}(\varphi, T)
\approx\frac{3T^4}{2\pi^2}  \int_{0}^{\infty} \d x x^2 \ln\left( 1 - e^{-\sqrt{x^2 +\bar{m}_{Z'}^2/T^2}}\right) \,.
\end{align}
We also include the thermal resummation to avoid breaking the perturbative expansion, and the complete form of the potential is given in the appendix. Near the field origin, the zero-temperature potential is flat, while $V_{T}(\varphi, T)\approx g_{B-L}^2T^2\varphi^2/2$, yielding a local minimum at $\varphi=0$, and generally leading to a supercooled FOPT that triggers the $U(1)_{B-L}$ and EW symmetries breaking~\cite{Witten:1980ez,Jinno:2016knw,Iso:2017uuu,Chiang:2017zbz,Marzo:2018nov,Bian:2019szo,Ellis:2019oqb,Kang:2020jeg,Ellis:2020nnr,Huang:2022vkf,Chun:2023ezg,Sagunski:2023ynd,Ahriche:2023jdq,Baldes:2023rqv,Salvio:2023ynn,Banerjee:2024fam,Liu:2024fly,Banerjee:2024cwv}.

\textbf{Baby universe formation}. Below the critical temperature $T_c$ (corresponding to time $t_c$), the effective potential $V_{\rm tot}(\varphi, T) $ possesses two local minima separated by a barrier, where the $\varphi=0$ one is the false vacuum, while the $\varphi\neq0$ one is the true vacuum. The universe then decays to the true vacuum with a rate per unit volume $\Gamma\approx T^4e^{-S_{3}(T)/T}$~\cite{Linde:1981zj}, and the volume fraction of false vacuum regions is given by~\cite{Guth:1981uk},
\bea\label{Ft}
&&F(t,t_d)=\Theta(t_d-t)+\Theta(t-t_d)\times\\
&&\quad\exp\left\{-\frac{4\pi}{3}\int_{t_d}^{t}\d t'\Gamma(t',t_d)a^3(t',t_d)\left[\int_{t'}^t\frac{v_w\d t''}{a(t'',t_d)}\right]^3\right\},\nn
\eea
where $\Theta$ is the Heaviside step function, $v_w$ is the expansion velocity of the bubble wall, and $t_d\geqslant t_c$ is the time when the bubble nucleation starts, which varies in different spatial positions. The notation $X(t, t_d)$ indicates a quantity $X$ evaluated at time $t$ in a region with nucleation time $t_d$.

During the FOPT, we consider two types of space regions: the normal Hubble patches with $t_d = t_c$, and delayed-decay regions with $t_d \gg t_c$. Due to the smallness of the vacuum decay rate and the randomness of the nucleation site, delayed regions must exist, and may further form a FVD surrounded by the true vacuum regions. Since the radiation energy is diluted by cosmic expansion, while the vacuum energy remains unchanged, the FVD forms a large energy density contrast and can collapse into a super-critical PBH, which contains a baby universe that connects with the parent universe via an untraversable wormhole~\cite{Blau:1986cw,Garriga:2015fdk}. We now calculate this process quantitatively. To simplify the notation, variables related to normal and delayed regions are distinguished by a subscript ``$\,d\,$''. For example, the scale factors are $a(t) \equiv a(t, t_c)$ for normal regions and $a_d(t) \equiv a(t, t_d)$ for delayed regions, and similarly for temperatures $T(t) \equiv T(t, t_c)$ and $T_d(t) \equiv T(t, t_d)$.

If a delayed region is still in the false vacuum until a time $t_i\in(t_c,t_d)$ when all surrounding regions have tunneled to the true vacuum, it constitutes a FVD. Under assumptions (i) the FVD is spherically symmetric and (ii) the thin-wall approximation can be applied to the boundary, we can employ Israel's junction condition~\cite{Israel:1966rt}, and obtain the simplified Einstein equation for the FVD boundary $\chi$~\cite{Tanahashi:2014sma, Deng:2020mds, Hashino:2025fse}
\begin{multline}\label{eq:EoM1}
\ddot{\chi}  + (4 - 3 a_d^2 \dot{\chi}^2) H_d \dot{\chi} + \frac{2}{a_d^2 \chi} ( 1 - a_d^2 \dot{\chi}^2) \\
= \left( \frac{3 \sigma}{4M_{\rm pl}^2} - \frac{\rho_{V}}{\sigma} \right) \frac{(1 - a_d^2 \dot{\chi}^2)^{3/2}}{a_d}\,,
\end{multline}
where $M_{\rm pl} = 2.435\times 10^{18}\,{\rm GeV}$, and $\rho_{V}=V_{\rm tot}(0, T_d) - V_{\rm tot}(\braket{\varphi}, T_d)$ is the vacuum energy in the FVD. The Hubble parameter $H_d(t)\equiv H(t,t_d)$ is given by the Friedmann equation $H_d^2  = (\dot{a}_d/a_d)^2 = (\rho_{V}+\rho_{R})/(3 M_{\rm pl}^2)$, where $\rho_{R}(t)$ is the radiation energy component, whose time evolution is determined by $\dot{\rho}_{R}+4H_d\rho_{R}=-\dot{\rho}_{V}$. In the thin-wall approximation, the surface energy is given by~\cite{Linde:1981zj}
\begin{align}
\sigma \approx \int_{0}^{\infty} \d r \left[ \frac{1}{2} \left( \frac{\d \hat\varphi}{\d r} \right)^2 + V_{\rm tot}(\hat\varphi, T_d(t_i)) \right] \,,
\end{align}
where $\hat\varphi(r)$ is the bounce solution. The initial conditions are $\chi(t_i)=\chi_i$ and $\dot\chi(t_i)=0$.

The physical radius of the FVD is $R(t) = a_d(t) \chi(t)$. A baby universe forms if, at the wormhole pinch-off time $t_b=t_i+t_H$, the physical radius exceeds the Hubble radius~\cite{Sato:1981gv,Garriga:2015fdk}: $R(t_b)>H_d^{-1}(t_b)$, where $t_H=H_d(t_i)R^2(t_i)/2$ is the free-fall time. In a region dominated by radiation, $R\propto\sqrt{t}$ grows slower than $H_d^{-1}\propto t$, preventing the inequality from being satisfied. However, in a region dominated by vacuum energy, $R$ grows exponentially while $H_d^{-1}$ is a constant, allowing the condition to be met. Therefore, the formation of a baby universe favors the FVD where the vacuum energy fraction is enhanced, as expected.

\textbf{Escaping from the eternal inflation}. Born from a FVD, the baby universe begins in the false vacuum $(h,\,\varphi)=(0,\,0)$, and is expanding exponentially due to vacuum energy domination. This is called secondary inflation or thermal inflation. In conventional theories, this inflation is problematic, as it occurs at low temperatures but is governed by high-scale physics that is decoupled from low-energy phenomena, making the inflation unstoppable. Consequently, such a thermal history is apparently in conflict with current cosmological observations, and hence, previously, we were not supposed to be living in a baby universe. However, in CC theories, due to the absence of tree-level mass terms, low-scale phase transitions could significantly affect the vacuum structure and terminate the secondary inflation.

As the baby universe cools to $T_{\rm QCD}\approx85~{\rm MeV}$, the QCD phase transition occurs. Induced by six-flavor massless quarks, this is a FOPT, different from the SM crossover~\cite{Pisarski:1983ms,Braun:2006jd,Guan:2024ccw}. The resulting chiral condensation induces a linear term $-y_th\ave{\bar t t}/\sqrt{2}$ in the Higgs potential via the top quark Yukawa coupling, leading to a Higgs VEV at $v_{\rm QCD}\approx100$ MeV. This VEV induces a negative mass squared term $-m_h^2v_{\rm QCD}^2\varphi^2/(4v_\varphi^2)$ along the $B-L$ direction. This term overwhelms the thermal mass term $g_{B-L}^2T^2\varphi^2/2$, eventually triggering the $U(1)_{B-L}$ and the EW symmetry breaking, terminating the inflation period, and starting the standard thermal cosmology~\cite{Witten:1980ez}. The huge vacuum energy stored in the false vacuum is then released to the space, reheating it to a high temperature $T_{\rm rh}\sim\rho_V^{1/4}$. As long as $T_{\rm rh}\gtrsim$ MeV, the big bang nucleosynthesis (BBN) is not affected and standard cosmology can be realized.

\textbf{The probability that we live in a baby universe}. The probability of a FVD being present during the supercooled FOPT can be factorized as $P_{\rm FVD}=P_{\rm FV}\times P_{\rm TV}$, where the former is the probability that a region remains in the false vacuum until $t_d$, while the latter is the probability that this region is surrounded by the true vacuum.  $P_{\rm FV}$ is calculated by integrating the probability that no bubbles nucleate within the prospective FVD volume up to time $t_d$,~\cite{Liu:2021svg,Kanemura:2024pae,Hashino:2025fse} i.e., 
\be\label{PFV}
P_{\rm FV}=\exp\left\{-\frac{4\pi}{3}\int_{t_c}^{t_d}\d t \Gamma_d(t) a_d^3(t)\chi^3(t)\right\}\,.
\ee
Since $P_{\rm FV}$ drops sharply with increasing $t_d$, and $t_d\geqslant t_b$ is required by the formation of the baby universe, we adopt $t_d=t_b$ to maximize $P_{\rm FV}$.

We now compute $P_{\rm TV}$. The false vacuum fraction in a normal patch is $F(t)$. We adopt a simplified picture: within an ensemble of such patches, a fraction $F(t)$ is in the false vacuum, while the remaining fraction $1-F(t)$ is in the true vacuum. Consider that the FVD at time $t_i$ is a sphere with a surface area $S_i=4\pi a_d^2(t_i)\chi^2(t_i)$. Modeling the surrounding space as a grid of cubic Hubble patches of side length $L_i=H_d^{-1}(t_i)$, the number of patches adjacent to the FVD surface is approximately $S_i/L_i^2$. For the FVD to be surrounded by true vacuum, all these patches must be in the true vacuum state. Therefore,
\be\label{PTV}
P_{\rm TV}=\left[1-F(t_i)\right]^{4\pi a_d^2(t_i)\chi^2(t_i)H_d^2(t_i)}\,.
\ee
This expression also explicitly requires $t_i>t_c$, as the true vacuum fraction must be nonzero.

Although $P_{\rm FVD}$ represents the probability that baby universes exist within a parent universe, it is not the probability that we reside in one. To calculate the latter, we consider the volume fraction of the would-be-baby-universe regions at $t_c$, when the parent universe is still homogeneous. The volume of a FVD at $t_c$ is $(4\pi/3)a_d^3(t_c)\chi^3(t_c)$, where we assume $\chi(t_c)=\chi_i$. At the same time, the volume of a normal patch is $(4\pi/3)H^{-3}(t_c)$. Consequently, the probability that we are inside a FVD is the probability that we live in a baby universe, which reads
\be\label{Pbaby}
P_{\rm baby}=\frac{P_{\rm FVD}a_d^3(t_c)\chi_i^3}{P_{\rm FVD}a_d^3(t_c)\chi_i^3+(1-P_{\rm FVD})H^{-3}(t_c)}\,.
\ee
For a given parameter set of the particle physics model, we vary $t_i$ and $\chi_i$ to obtain a maximal $P_{\rm baby}$. We emphasize that this resulting $P_{\rm baby}$ specifically describes the probability that we inhabit a baby universe within the context of this minimal $B-L$ model; it is not the probability that the entire scenario is realized in nature.

\begin{figure}[t]
\centering
\includegraphics[width=0.95\linewidth]{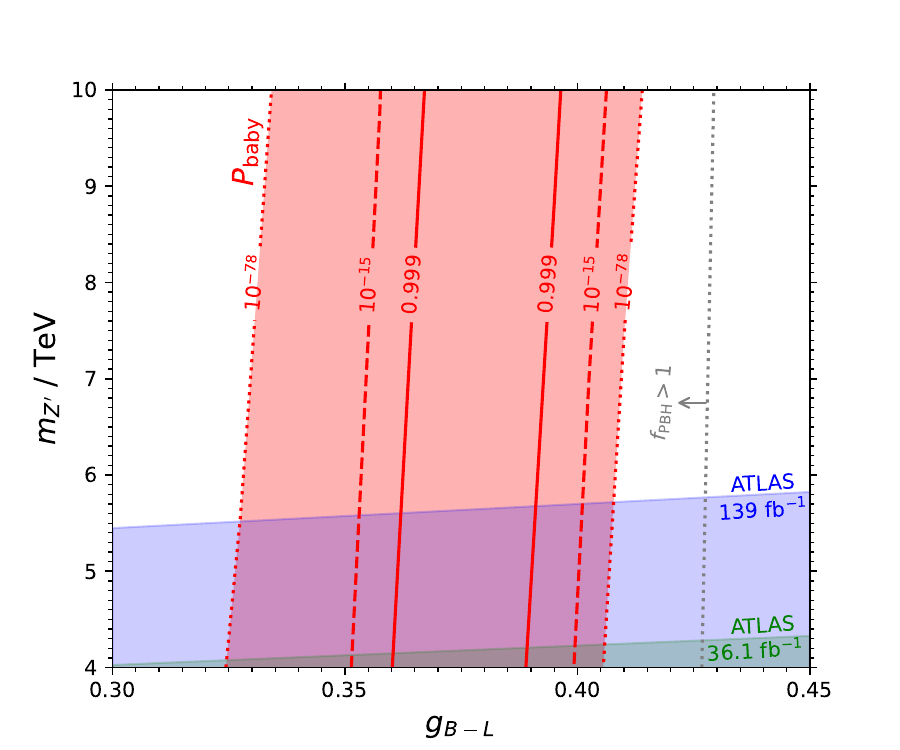}
\caption{Numerical results of the minimal $B-L$ model. The red solid lines represent contours of $P_{\rm baby}$. If we assume no extra dilution mechanism, $f_{\rm PBH}=\Omega_{\rm PBH}/\Omega_{\rm DM}$ in the parent universe is larger than 1 in the left part of the black dashed line. The green (blue) shaded region is the constraint from the ATLAS dilepton search with the integrated luminosity of $36.1~{\rm fb}^{-1}$~\cite{ATLAS:2017fih} ($139~{\rm fb}^{-1}$~\cite{Queiroz:2024ipo}).
}
\label{fig:2dimplot}
\end{figure}

\textbf{Numerical results}. Using the \texttt{CosmoTransitions}~\cite{Wainwright:2011kj} package to obtain the vacuum decay rate $\Gamma(T)$, we resolve the FOPT dynamics and the baby universe formation process in the minimal $B-L$ model, adopting $v_w=1$. The resulting contours of $P_{\rm baby}$ are presented in Fig.~\ref{fig:2dimplot} as red lines. We can see that $P_{\rm baby}$ can be as large as $>0.999$ within $0.36\lesssim g_{B-L}\lesssim0.39$, but it decreases very rapidly when $g_{B-L}$ is out of this range. The sensitive dependence of $P_{\rm baby}$ on $g_{B-L}$ is from the bounce action $S_3/T \propto g_{B-L}^{-3}$~\cite{Iso:2017uuu}, which determines the vacuum decay rate on the exponent, i.e., $\Gamma(T)\propto e^{-S_3/T}$. A small $g_{B-L}$ suppresses $\Gamma(T)$, thus enhancing supercooling and favoring baby universe formation. However, an excessively small $g_{B-L}$ also causes normal patches to undergo a supercooled FOPT, preventing the existence of the required true vacuum regions that surround the FVDs and thus reducing $P_{\rm baby}$.

The lowest value of $P_{\rm baby}$ shown in the contours is $10^{-78}$, based on a rough estimate based on inflationary cosmology. An e-folding number of $N \gtrsim 60$ is needed to explain the initial condition problems of the universe~\cite{Kolb:1990vq}. This means that during the first inflation period, the parent universe becomes at least $e^{N}$ times larger than its size before inflation. As a result, the minimal total number of Hubble patches in our universe is roughly given by $N_{\rm universe} \gtrsim (e^{N})^3 = e^{180} \approx 10^{78}$. Therefore, if $P_{\rm baby}\gtrsim10^{-78}$, at least one baby universe could be formed in the parent universe, and there is a possibility that we live in a baby universe.

Given a set of $g_{B-L}$ and $m_{Z'}$, we can derive the PBH profile of the parent universe. For $g_{B-L}\lesssim 0.43$, the predicted PBH abundance exceeds the observed dark matter (DM) density, i.e., $f_{\rm PBH} = \Omega_{\rm PBH}/\Omega_{\rm DM} > 1$. While this discrepancy might suggest that we are living in a baby universe, this conclusion must be viewed with caution. The evolution of cosmic history after PBH formation can substantially modify the PBH abundance observed today. Since our model does not include a DM candidate or explain the baryon asymmetry, additional mechanisms are required to address these issues. Those late-time processes could dilute the PBH abundance, potentially yielding $f_{\rm PBH} \lesssim 1$ in the parent universe. Therefore, the line indicating $f_{\rm PBH} = 1$ is presented as a dashed black line in Fig.~\ref{fig:2dimplot} for illustrative purposes, rather than a definitive prediction.

Although the baby universe formation mechanism itself imposes no intrinsic upper bound on the $Z'$ mass, considerations of the naturalness of the EW scale require $m_{Z'}$ at the TeV scale~\cite{Iso:2009nw}. Consequently, we restrict our parameter space to $m_{Z'}\leqslant10$ TeV, a region that is largely accessible to collider experiments. The existing LHC bounds~\cite{ATLAS:2017fih, Queiroz:2024ipo} are shown as the blue and green regions in Fig.~\ref{fig:2dimplot}. The High-Luminosity LHC with $\sqrt{s}=14\,{\rm TeV}$ and the integrated luminosity $3\,{\rm ab}^{-1}$ is expected to probe $m_{Z^{\prime}} < 7.578~{\rm TeV}$ via dilepton searches with $g_{B-L} = 0.5$~\cite{Queiroz:2024ipo}. Furthermore, future colliders like the FCC-hh with $\sqrt{s} =100~{\rm TeV}$ and integrated luminosity $30~{\rm ab}^{-1}$ could extend this reach up to $m_{Z'}=40~{\rm TeV}$~\cite{Liu:2022kid}, providing an efficient probe for our scenario.

The gravitational wave (GW) provides another probe. Since the super-horizon FVD becomes the baby universe, it is expected that the GWs produced by the FOPTs in the parent universe cannot propagate into the baby universe. Inside the baby universe, the $U(1)_{B-L}$ FOPT triggered by low-scale QCD transition is extremely strong; yet its duration is typically too short to produce significant GWs~\cite{Liu:2024fly}. However, the absence of GWs can itself provide valuable information for our scenario. In Fig.~\ref{fig:2dimplot}, for parameters yielding $P_{\rm baby}>0.999$, there would be a relatively large GW signal in the parent universe. For instance, the signal-to-noise ratio at the future LISA detector~\cite{LISA:2017pwj} is projected to be $\sim10^3-10^4$ using the numerical formulae~\cite{Caprini:2019egz}, assuming a data-taking duration of $0.75\times4$ years. This leads to a distinctive and testable prediction: if future collider experiments measure $0.36\lesssim g_{B-L}\lesssim0.39$, while next-generation space-based GW detectors (e.g., LISA~\cite{LISA:2017pwj}, TianQin~\cite{TianQin:2020hid}, Taiji~\cite{Ruan:2018tsw}, or DECIGO~\cite{Kawamura:2011zz}) observe no corresponding stochastic FOPT GWs, it would constitute strong evidence that we are living in a baby universe within this framework.

We also comment on problems with DM and baryon asymmetry. The secondary inflation in the baby universe ends at a very low temperature $T_*\lesssim T_{\rm QCD}$, while subsequent reheating raises the temperature to $T_{\rm rh}\sim $ TeV in the parameter space of interest. This entropy production dilutes any preexisting DM and baryon asymmetry inherited from the parent universe by a factor of $(T_*/T_{\rm rh})^3\lesssim10^{-12}$. However, they can be produced in various ways after reheating. For $T_{\rm rh}\gtrsim100\,{\rm GeV}$, standard mechanisms become viable: DM could consist of weakly interacting massive particles produced via freeze-out, while the baryon asymmetry could originate from EW baryogenesis or low-scale leptogenesis. If $T_{\rm rh}$ is lower, non-standard mechanisms such as supercooled dark matter~\cite{Hambye:2018qjv, Wong:2023qon} or baryogenesis via meson decays~\cite{Alonso-Alvarez:2019fym,Nelson:2019fln} would be required.

\textbf{Conclusion}. We proposed that we could live in a baby universe that is produced during a supercooled FOPT of the parent universe. The eternal secondary inflation in conventional baby universe scenarios, which ruins any possibility that we are living in such universes, is prevented by the low-scale chiral symmetry breaking in the CC models. Building upon this concept, we define the variable $P_{\rm baby}$ to quantify the probability that we reside in a baby universe, marking a novel step in the study of the multiverse. A concrete minimal $B-L$ model is provided to illustrate this idea, which could yield a $P_{\rm baby}>0.999$ in a considerable portion of the parameter space, and be tested by combining collider and GW experiments. Our work extends the study of the multiverse, opening up a novel possibility of the universe's origin. As the CC principle is motivated by the SM hierarchy problem, our work suggests a speculative but intriguing deep relation between the origin of the universe and the quantum corrections to the elementary scalar boson mass.

\textbf{Acknowledgments}. M.T. would like to thank Chul-Moon Yoo and Takumi Kuwahara for helpful and fruitful discussions. The authors gratefully acknowledge the valuable discussions and insights provided by the members of the Collaboration on Precision Tests and New Physics (CPTNP). The work is partly supported by the National Natural Science Foundation of China under Grant Nos. 12235001 and 12305108.

\bibliographystyle{apsrev}
\bibliography{references} 

\newpage

\onecolumngrid

\appendix
\section{Details of the one-loop potential \label{appendixA}}

In this appendix, we explain the derivation of the effective potential in our model. The one-loop correction to the scalar potential at zero temperature is given by~\cite{Chiang:2017zbz}
\begin{multline}\label{eq:V1l0}
V_{1\ell}(\varphi) = \frac{\overline{m}_{\varphi}^4}{64\pi^2} \left( \ln \frac{\overline{m}_{\varphi}^2}{\mu^2} - \frac{3}{2} \right) 
+  \frac{3\overline{m}_{Z'}^4}{64\pi^2} \left( \ln \frac{\overline{m}_{Z'}^2}{\mu^2} - \frac{5}{6} \right)
+  \frac{\overline{m}_{G,\xi}^4}{64\pi^2} \left( \ln \frac{\overline{m}_{G,\xi}^2}{\mu^2} - \frac{3}{2} \right) - \frac{( \xi \overline{m}_{Z'}^2)^2}{64\pi^2} \left( \ln \frac{\xi \overline{m}_{Z'}^2}{\mu^2} - \frac{3}{2} \right), 
\end{multline}
where the $R_{\xi}$ gauge and $\overline{\text{MS}}$ scheme are used, $\mu$ represents the renormalization scale, and the field-dependent masses read $\overline{m}_{\varphi}^2 = 3 \lambda_{\varphi} \varphi^2$, $\overline{m}_{Z'}^2 = 4 g_{B-L}^{2} \varphi^2$, $\overline{m}_{G, \xi}^2 = \overline{m}_{G}^2 + \xi \, \overline{m}_{Z'}^2$, and $\overline{m}_{G}^2 = \lambda_{\varphi} \varphi^2$.

We impose the tadpole condition on the tree-level part of the potential, $\partial V_{\rm tree}/\partial \varphi|_{\varphi = v_{\varphi}} = 0$, where $\braket{\varphi} = v_{\varphi}/\sqrt{2}$. This condition gives
\begin{align}
\label{eq:tadpole}
\lambda_{\varphi} \approx - \frac{3 m_{Z'}^4}{16 \pi^2 v_{\varphi}^4} \left( \ln \frac{m_{Z'}^2}{\mu^2} - \frac{1}{3} \right) \,,
\end{align}
where masses without the bar such as $m_{Z'}$ means physical masses. Since the contribution from the second term in Eq.~\eqref{eq:V1l0} gives higher-order effects, we can neglect it in deriving Eq.~\eqref{eq:tadpole}. 
Substituting Eq.\,\eqref{eq:tadpole} into Eq.~\eqref{eq:V1l0} and neglecting higher order terms (including $\lambda_{\varphi}$), we obtain
\begin{align}
V_{1\ell}(\varphi) \approx \frac{3 \overline{m}_{Z'}^4}{64 \pi^2} \left( \ln \frac{\varphi^2}{v_{\varphi}^2} - \frac{1}{2} \right) \,.
\end{align}
We note that the gauge-dependent part can be canceled out up to one-loop level at zero temperature.

The thermal correction part is given by
\begin{align}
V_{T}(\varphi, T)
= 3 \frac{T^4}{2\pi^2} I_{B} \left( \frac{\overline{m}_{Z'}^2}{T^2} \right) \,. 
\end{align}
where 
\begin{align}
I_{B}(a) = \int_{0}^{\infty} \d x x^2 \ln \left[ 1 - e^{-\sqrt{x^2 + a}} \right] \,. 
\end{align}
Since the thermal correction from the field $\Phi$ corresponds to a two-loop contribution, we ignore it. 

In addition, we have to take into account the thermal resummation to avoid breaking down the perturbative expansion. 
In this case, field-dependent masses for bosonic fields in $V_{1\ell}$ and $V_{T}$ should be replaced with the corresponding thermal masses. 
As a result, the total finite temperature potential can be given by
\begin{align}
\begin{aligned}
V_{\rm eff}(\varphi, T) =
& - 3 \frac{\overline{m}_{Z'}^4}{64\pi^2} \left( \ln \frac{m_{Z'}^2}{\mu^2} - \frac{1}{3} \right)
+ \frac{\overline{M}_{L}^4}{64 \pi^2} \left( \ln \frac{\overline{M}_{L}^2}{\mu^2} - \frac{3}{2} \right) 
+ 2 \frac{\overline{M}_{T}^4}{64 \pi^2} \left( \ln \frac{\overline{M}_{T}^2}{\mu^2} - \frac{1}{2} \right) \\
& + \frac{(\xi \overline{m}_{Z'}^2 + \Delta m_{\varphi}^2)^2 }{64\pi^2} \left( \ln \frac{\xi \overline{m}_{Z'}^2 + \Delta m_{\varphi}^2}{\mu^2} - \frac{3}{2} \right)  - \frac{(\xi \overline{m}_{Z'}^2)^2 }{64\pi^2} \left( \ln \frac{\xi \overline{m}_{Z'}^2 }{\mu^2} - \frac{3}{2} \right) \\
& + \frac{T^4}{2\pi^2} \left[ I_{B} \left( \frac{\overline{M}_{L}^2}{T^2} \right) + 2 I_{B} \left( \frac{\overline{M}_{T}^2}{T^2}  \right) + I_{B} \left( \frac{\xi \overline{m}_{Z'}^2 + \Delta m_{\varphi}^2}{T^2} \right) - I_{B} \left( \frac{\xi \overline{m}_{Z'}^2 }{T^2} \right) \right] \,,
\end{aligned}
\end{align}
with
\begin{align}
\begin{aligned}
\overline{M}_{L}^2 = \overline{m}_{Z'}^2 + \frac{4}{3} g_{B-L}^2 T^2 \,, \quad \overline{M}_{T}^2 = \overline{m}_{Z'}^2 \,, \quad \Delta m_{\varphi}^2 = g_{B-L}^2 T^2 \,. 
\end{aligned}
\end{align}
It should be noted that the finite temperature potential depends on the gauge parameter $\xi$ due to the thermal resummation effect~\cite{Chiang:2017zbz}. The renormalization scale $\mu$ is taken to be $\mu = v_{\varphi}$, and $\xi=0$ (Landau gauge) is adopted in the calculation.

\twocolumngrid

\end{document}